\documentclass[eqsecnum,
aps,
11pt,%
final,
nobibnotes,%
nofootinbib,
noshowpacs,%
centertags]%
{revtex4}
\usepackage{stmaryrd}
\usepackage{amsmath}    
\usepackage{amssymb}
\begin{document}

%
\title[Generalized Wigner-In\"on\"u Contractions]{Generalized Wigner-In\"on\"u Contractions and Maxwell (Super)Algebras}

\author{Jerzy Lukierski}

\email{lukier@ift.uni.wroc.pl}
\affiliation{Institute for Theoretical Physics, University of Wroc\l aw \\
pl. Maxa Borna 9, 50-204 Wroc\l aw}

\begin{abstract}
We consider a class of generalized In\"on\"u-Wigner contraction for semidirect product of two particularly related semisimple Lie (super)algebras. The special class of such contractions provides  D=4 Maxwell algebra and recently introduced simple D=4  Maxwell superalgebra. Further we present two types of D=4 $N$-extended Maxwell superalgebras,
 the nonstandard one  for any $N$
 with $\frac{1}{2}N(N-1)$ central charges and 
 the standard one,  for even N=2k, with 
  $k(2k-1)$ internal symmetry  generators.
\end{abstract}
\maketitle
\section{Introduction}
Standard Wigner-In\"on\"u contraction (IW) \cite{mosluk1,mosluk2} one applies usually to the symmetric cosets of simple Lie algebras, i.e. $\widehat{g}=\widehat{h}\oplus \widehat{k}$, where
\begin{equation}\label{wz1.1}
[\widehat{h},\widehat{h}] \subset \widehat{h} \qquad
[\widehat{h}, \widehat{k} ] \subset \widehat{k} \qquad
[\widehat{k},\widehat{k}] \subset \widehat{h} \, .
\end{equation}
After the following rescaling of the generators $h_\alpha \in \widehat{h}$ and $k_i \in \widehat{k}$
\begin{equation}\label{wz1.2}
h_\alpha = h'_\alpha \qquad k_i = \lambda k'_i \, ,
\end{equation}
one obtains in the limit $\lambda \to \infty$ the contracted algebra $\widehat{g}{}'$ which becomes a semidirect product
\begin{equation}\label{wz1.3}
\widehat{g}{}' = \widehat{h} 
 \ltimes
  \widehat{k} {}' \,,
\end{equation}
with Abelian sector ${\widehat{k}}{}'$. In such a way one can obtain\footnote{In this note
 for simplicity we shall consider  only the case D=4.} e.g. the Poincar\'{e} algebra $O(3,1) \supsetplus T^4$ from anti-de-Sitter ($\widehat{g} = O(3,2)$) 
or de-Sitter $(\widehat{g}=O(4,1))$
algebras, as well as the simple superPoincar\'{e} algebra $\widetilde{\cal P}_4= O(3,1) \supsetplus (T^4 \oplus Q^4)$ ($Q^4$ describe four real odd supercharges) by choosing $\widehat{g}=OSp(1|4)$ (N=1, D=4 $AdS$ superalgebra) and $\widehat{h}=
O(3,1)$ (D=4 Lorentz algebra). It should be mentioned that the standard W-I contraction procedure has been extended as well to quantum - deformed Lie algebras \cite{mosluk3,mosluk4} and provided first in the literature quantum Poincar\'{e} algebra \cite{mosluk5,mosluk6}.

The standard W-I contraction scheme can be extended in various ways to the non simple Lie algebras (see e.g. \cite{mosluk7}). The general scheme which we shall use below was proposed by Weimar-Woods \cite{mosluk8} and discussed further in \cite{mosluk9}.\footnote{We shall call it the 
WI~W-W (Wigner, In\"on\"u, Weimar-Woods) contraction scheme.}

Let us assume that the nonsimple Lie algebra $\widehat{G}$ is decomposed into the n+1 sets of generators $\widehat{g}^{(k)}$ ($k=0,1,\ldots , n$)
\begin{equation}\label{wz1.4}
\widehat{G} = \widehat{g}^{(0)} \oplus \widehat{g}^{(1)} \oplus \ldots \oplus \widehat{g}^{(n)} \,,
\end{equation}
where the following conditions are satisfied \cite{mosluk8}
\begin{equation}\label{wz1.5}
[ \widehat{g}^{(k)} , \widehat{g}^{(l)} ] 
\subset 
{\displaystyle{\mathop{\bigoplus}\limits_{s\leq k+l}}} \
\widehat{g}^{(s)}
\end{equation}

The contraction of (\ref{wz1.4}) is obtained by properly rescaling the generators
\begin{equation}\label{wz1.6}
\widehat{g}^{(i)} = \lambda ^{a_i} \, \widehat{g}^{(i)} \, ,
\end{equation}
with the choice of $a_i$ providing the finite limits of the algebra (\ref{wz1.5}) if $\lambda \to \infty$.

Clearly it follows that 

i) ${\widehat{g}}^{(0)}$ describes the subalgebra of $\widehat{G}$

ii) if n=1 we obtain standard decomposition of Lie algebra into a Riemannian pair with
${\widehat{g}}^{(1)}$ describing a coset ${\widehat{k}}$ (see (\ref{wz1.1}))

In this paper we shall consider the contraction of the following  semidirect product which provides a particular choice of (\ref{wz1.4}--\ref{wz1.5}) for n=2\ \footnote{The extension of presented scheme to superalgebras $\widehat{G}$ is straightforward.}
\begin{equation}\label{wz1.7}
\widehat{G} = \widehat{h} 
\ltimes
 \widehat{g} = \widehat{h} 
 \ltimes
  (\widehat{h} \oplus \widehat{k}) \,.
\end{equation}
We see that the algebra $\widehat{h}$ occurs twice and all algebraic relations in algebra (\ref{wz1.7}) are already given by (\ref{wz1.1}). One can
rewrite (\ref{wz1.7}) using rescaled
 generators.

\begin{equation}\label{wz1.8}
\widehat{G} = \widehat{h} 
\ltimes
 (\widehat{h}{ }' \oplus \widehat{k}{ }'') \,.
\end{equation}
where
\begin{equation}\label{wz1.9}
{h}_\alpha = \lambda^2 {h}'_{\alpha}
\qquad
 k_i = \lambda k''_i \,.
\end{equation}
In the limit $\lambda \to \infty$ we obtain the following set of commutation relations
\begin{equation}\label{wz1.10}
\begin{array}{lll}
[\widehat{h}, \widehat{h}] \subset \widehat{h} \quad
&
[\widehat{h},\widehat{h}{ }'] \subset \widehat{h}{ }' 
\quad
& [\widehat{h}, \widehat{k} { }''] \subset k''
\\[5pt]
[\widehat{k}{ }'', \widehat{k}{ }''] \subset \widehat{h}{ }' \quad
&
[\widehat{h} { }',\widehat{k}{ }''] = 0
\quad
&
 [\widehat{h} { }', \widehat{h} { }'] = 0 \,.
\end{array}
\end{equation}
We see that the generator $k''_i$ are the ``Lie-algebraic roots'' of the Abelian generators
$h'_\alpha$.\footnote{This terminology is used in analogy with the statement that supercharges in semisimple Lie superalgebras  are called the ``SUSY roots'' of its bosonic generators.}

In this note we shall apply such a contraction scheme to the derivation in Sect.~2 of Maxwell algebra \cite{mosluk10}--\cite{mosluk12} and simple nonstandard \cite{mosluk13} as well as    standard \cite{mosluk14} Maxwell superalgebras. In Sect.~3 by performing suitable 
WI~W-W contractions we shall obtain
new results: 
 the  extended nonstandard and extended standard Maxwell superalgebras.

In Sect.~4 we shall recall briefly the geometric interpretation of Maxwell (super)symmetries    \cite{mosluk15,mosluk14} and present final comments.

\section{Maxwell Algebra and Simple Maxwell Superalgebras by WI~W-W Contraction Scheme}

{\textbf {i) Maxwell algebra (arbitrary D) }}

Let us choose in the formula (\ref{wz1.7})
\begin{equation}\label{wz2.1}
\widehat{h}=O(D-1,1) \qquad \widehat{g}=O(D-1,2) \,.
\end{equation}
We denote the generators $\widehat{h}$ of Lorentz algebra by ${{M}}_{\mu\nu}$, and the generators
$\widehat{g}$ of $D$-dimensional $AdS$ algebra by ($\widetilde{M}_{\mu\nu}, {\cal P}_{\mu}$)
(we put $AdS$ radius $R=1$). One gets
\begin{equation}
\begin{split}
\label{wz2.2}
&[M_{\mu\nu}, M_{\rho\tau} ] =
i (\eta_{\nu[\rho} \, M_{\mu]\tau} -
\eta_{\mu[\rho}\, M_{\nu]\tau})
\\[5pt]
&[M_{\mu\nu}, {\widetilde{M}}_{\rho\tau} ] =
i (\eta_{\nu[\rho} \, {\widetilde{M}}_{\mu]\tau} -
\eta_{\mu[\rho}\, {\widetilde{M}}_{\nu]\tau})
\\[5pt]
&[{\widetilde{M}}_{\mu\nu}, {\widetilde{M}}_{\rho\tau} ] =
i (\eta_{\nu[\rho} \, {\widetilde{M}}_{\mu]\tau} -
\eta_{\mu[\rho}\, {\widetilde{M}}_{\nu]\tau})
\\[5pt]
&[{M}_{\mu\nu}, {\cal P}_{\rho} ] =
i (\eta_{\nu\rho} \, {\cal P}_{\nu} -
\eta_{\mu\rho}\, {\cal P}_{\mu})
\\[5pt]
&[{\widetilde{M}}_{\mu\nu}, {\cal P}_{\rho} ] =
i (\eta_{\nu\rho} \, {\cal P}_{\mu} -
\eta_{\mu\rho}\,{\cal P}_{\nu})
\\[5pt]
&[{\cal P}_{\mu} , {\cal P}_{\nu} ] = i {\widetilde{M}}_{\mu\nu}
\end{split}
\end{equation}

The rescaling ($M_{\mu\nu}$ remain unchanged)
\begin{equation}\label{wz2.3}
 \widetilde{M}_{\mu\nu} = \lambda^2 Z_{\mu\nu}
\qquad
{{\cal P}}_{\mu} = \lambda \, P_\mu
\end{equation}
provides in the limit $\lambda\to \infty$ the Maxwell algebra:

\begin{subequations}
\begin{align}\label{wz2.4a}
&[M_{\mu\nu}, M_{\rho\tau} ] =
i (\eta_{\nu[\rho} \, M_{\mu]\tau} -
\eta_{\mu[\rho}\, M_{\nu]\tau})
\\
\label{wz2.4b}
&[M_{\mu\nu}, Z_{\rho\tau} ] =
i (\eta_{\nu[\rho} \, Z_{\mu]\tau} -
\eta_{\mu[\rho}\, Z_{\nu]\tau})
\\
\label{wz2.4c}
&[{{M}}_{\mu\nu}, {{P}}_{\rho} ] =
i (\eta_{\nu\rho} \, {{P}}_{\mu} -
\eta_{\mu\rho}\, P_{\nu})
\\
\label{wz2.4d}
&[{P}_{\mu}, { P}_{\nu} ] =
i  \, Z_{\mu\nu}
\\
\label{wz2.4e}
&[{{Z}}_{\mu\nu}, {Z}_{\rho\tau} ] =
[P_{\mu}, Z_{\rho, \tau}] =0 \,.
\end{align}
\end{subequations}
We see from (\ref{wz2.4b},\ref{wz2.4d},\ref{wz2.4e}) that the Abelian generators $Z_{\rho\tau}$
can be called tensorial central charges\footnote{This may explain why Maxwell algebra was rediscovered in \cite{mosluk17} under the name of tensorial extension of Poincar\'{e} algebra.}.
We add that one can obtain the Maxwell algebra  as well if we choose in (\ref{wz2.1}) that
$\widehat{g} = O(D,1)$ ($D$-dimensional de-Sitter) algebra \cite{mosluk16}.

{\textbf{ii) Simple nonstandard D=4 Maxwell  superalgebra}}

The structure presented by formulae (\ref{wz1.7}--\ref{wz1.10}) can be obtained as well if 
$\widehat{g}$ in (\ref{wz1.7}) describes the semisimple Lie superalgebra \cite{mosluk8}. 
We choose in (\ref{wz1.7}) for D=4
\begin{equation}\label{wz2.5}
\widehat{h} = O(3,1) \qquad \widehat{g} =OSp(1|4) \,.
\end{equation}
The $OSp(1|4)$ relations extend the bosonic sector $Sp(4;R)=O(3,2)=({\widetilde{M}}_{\mu\nu}, {\cal P}_\mu)$ by four real supercharges ${\widetilde{Q}}_A \, (A=1\ldots 4)$ as follows:\footnote{We use real Majorana realization of D=4 Dirac algebra, with 
$C=\gamma_0$ ($C^T=-C,\ C^2=-1$).}
\begin{subequations}
\begin{equation}
\begin{split}
\label{wz2.6a}
&
\{
{\widetilde{Q}}_A , {\widetilde{Q}}_B\}
= \frac{1}{2}
(C\, \gamma^\mu )_ {AB}
{\cal P}_\mu 
+
\frac{1}{2}
(C\, \sigma^{\mu\nu})_{AB} {\widetilde{M}}_{\mu\nu}
\\[5pt]
&[
{\widetilde{M}}_{\mu\nu}, {\widetilde{Q}}_{A}
]
= \frac{1}{2} (\sigma_{\mu\nu})_{AB}
{\widetilde{Q}}^{B}
\\[5pt]
&[
{\cal P}_{\mu}, {\widetilde{Q}}_{A}
]
= \frac{1}{2} (\gamma_{\mu})_{AB}
{\widetilde{Q}}^{B}
\end{split}
\end{equation}
Besides we have the relations
\begin{equation}
\label{wz2.6b}
[
{{M}}_{\mu\nu}, {\widetilde{Q}}_{A}
]
= \frac{1}{2} (\sigma_{\mu\nu})_{AB}
{\widetilde{Q}}^{B}
\end{equation}
\end{subequations}
where $M_{\mu\nu} \in \widehat{h}$ and $\widetilde{M}_{\mu\nu} \in \widehat{g}$
 (see (\ref{wz2.5})).

If we keep the rescaling (\ref{wz2.3}) of  bosonic (even) generators the only 
 rescaling of odd generators 
 leading in the limit $\lambda \to \infty $ to   a nontrivial anticommutator of supercharges is provided by
\begin{equation}\label{wz2.7}
{\widetilde{Q}}_A = \lambda \, Q_A \,.
\end{equation}
In such a case we obtain as a contraction limit
 $\lambda \to \infty $
\begin{equation}\label{wz2.8}
\{
Q_A, Q_B \} =
\frac{1}{2}\, 
 C(\sigma^{\mu\nu})_{AB} \, Z_{\mu\nu}
\end{equation}
and
\begin{equation}\label{wz2.9}
[Q_A, Z_{\mu\nu} ] = [Q_A , {\cal P}_\mu ] =0
\end{equation}

Such supersymmetrization of Maxwell algebra was introduced by Soroka and Soroka 
\cite{mosluk13,mosluk17}. In such nonstandard supersymmetrization scheme only the tensorial central charges are supersymmetrized, and the fourmomentum generators, including energy, are not expressed as bilinear expressions in terms of supercharges $Q_A$.

Concluding, the N=1 $AdS$ superalgebra can provide by contraction only the nonstandard
supersymmetrization of Maxwell algebra.
 In such a framework, contrary to the scheme of SUSY quantum mechanics the Hamiltonian is not expressed in terms of supercharges.

{\textbf{iii) Standard Maxwell superalgebra by WI~W-W contraction of $\boldmath \boldsymbol OSp(2|4)$.}}

Let us write down the basic relation of N=2 $AdS$ superalgebra ($r,s=1,2$) 
\begin{equation}\label{wz2.10}
\{
{\widetilde{Q}}^r_A , {\widetilde{Q}}^s_B 
\}
=
\frac{1}{2} \, \delta^{rs} 
[(C \, \sigma^\mu)_{AB} {\cal P}_\mu 
+
(C \, \sigma^{\mu\nu})_{AB} \,{\widetilde{M}}_{\mu\nu}
+
 \in^{rs} \, C_{AB} \, Z \, ,
\end{equation}
where internal $O(2)$ symmetry generator $Z$ rotates a pair of multiplets of supercharges
\begin{equation}\label{wz2.11}
[{\widetilde{Q}}^r_A, Z] = 
\in^{rs} \, {\widetilde{Q}}^{s}_A \, .
\end{equation}

In order to supersymmetrize after contraction  both generators $P_\mu$ and $Z_{\mu\nu}$ we shall  rewrite the 
superalgebra (\ref{wz2.9}) in terms of suitably projected supercharges
\begin{equation}\label{wz2.12}
{\widetilde{Q}}^{r(\pm)}_A  \equiv
{\widetilde{Q}}^r_A \pm
\in^{rs}
(\gamma_5)_{AB}\, {\widetilde{Q}}^{s}_B \,.
\end{equation}
One gets the following basic superalgebraic relations
\begin{equation}\label{wz2.13}
\begin{split}
\{
{\widetilde{Q}}^{r(\pm)}_A,
{\widetilde{Q}}^{s(\pm)}_B
\}
&
=\frac{1}{2} P^{(+)rs}_{\quad AC}
(C \,\sigma^{\mu\nu})_{CB} \, {\cal P}_{\mu}
\\[5pt]
\{
{\widetilde{Q}}^{r(+)}_A,
{\widetilde{Q}}^{s(-)}_B
\}
&
=\frac{1}{2} P^{(+)rs}_{\quad AC}
(C \,\sigma^{\mu\nu})_{CB} \, {M}_{\mu\nu}  
\\[5pt]
& \ 
+ P^{(+)rt}_{\quad AC} \, \in^{ts} \, C_{CB} \cdot Z \,.
\end{split}
\end{equation}
By choosing besides (\ref{wz2.3}) the rescaling
 \cite{mosluk19}

\begin{equation}\label{wz2.14}
\begin{split}
{\widetilde{Q}}_A^{r(+){} \prime} 
= \lambda^{1/2} \, Q^{r(+)}_A
\\[5pt]
{\widetilde{Q}}^{r(-){ }\prime } _A
= \lambda^{3/2} \, Q^{r(-)}
\end{split}
\end{equation}
and additionally
\begin{equation}\label{wz2.15}
Z = \lambda^2 \, B_5
\end{equation}
one gets the following $\lambda \to \infty$ contraction limit
of the relations (\ref{wz2.12})
\begin{equation}\label{wz2.16}
\begin{split}
\{
Q^{r(+)} _A , Q^{s(-)}_B 
\}
&
= \frac{1}{2} P^{(\pm)rs}_{\quad AC} (C \, \gamma^\mu)_{CB}
{\cal P}_\mu
\\[5pt]
\{
Q^{r(+)} _A , Q^{s(-)}_B 
\}
&
= \frac{1}{2} P^{(+)rs}_{\quad AC} (C \, \sigma^{\mu\nu})_{CB}
{Z}_{\mu\nu}
\\[5pt]
&
\ 
+ \, P^{(+)rt}_{\quad AC} \, \in^{ts}\, C_{CB} \, B_5\,.
\end{split}
\end{equation}
In such a way we obtain the standard Maxwell superalgebra centrally extended by the Abelian generator $B_5$ as obtained 
recently in \cite{mosluk14,mosluk19}\footnote{For complete set of superalgebraic relations, extending odd-odd sector (\ref{wz2.16}) by odd-even and even-even ones see \cite{mosluk14,mosluk19}.}. It should be added that in \cite{mosluk14} the projected components of supercharges were expressed as two-component complex Weyl spinors, and the technique of projected supercharges was not used.

\section{The extended D=4 Maxwell Superalgebras}

The $N$-extended $AdS$ superalgebra $OSp(4|N)$ is given by the following basic relation satisfied by $N$ multiplets of supercharges $Q^r_A$ ($r=1\ldots N$)
\begin{equation}\label{wz3.1}
\{
{\widetilde{Q}}^r_A , {\widetilde{Q}}^s_B 
\}
= \frac{1}{2} 
\delta^{rs} 
[(C\, \gamma^\mu)_{AB} {\cal P}_\mu 
+ (C \, \sigma^{\mu\nu})_{AB} \, {\widetilde{M}}_{\mu\nu}]
+ {\widetilde{T}}^{rs} \, C_{AB} \,,
\end{equation}
where ${\widetilde{T}}^{rs} = - {\widetilde{T}}^{sr}$ describe
$\frac{N(N-1)}{2}\,$ internal $O(N)$  symmetry generators
\begin{equation}\label{wz3.2}
\begin{split}
[{\widetilde{T}}^{rs}, {\widetilde{T}}^{tu} ]
&
=
\delta^{rt} {\widetilde{T}}^{su}
- \delta^{st}
{\widetilde{T}}^{ru}
\\[5pt]
&\ +\, \delta^{ru}
{\widetilde{T}}^{st} +
\delta^{su}
{\widetilde{T}}^{rt}\,.
\end{split}
\end{equation}
They act on supercharges ${\widetilde{Q}}^{r}_A$ as follows
\begin{equation}\label{wz3.3}
[
{\widetilde{T}}^{rs}, {\widetilde{Q}}^{t}_A
]
= \tau^{rs;tu}\, {\widetilde{Q}}^u_A \,,
\end{equation}
where $\tau^{rs;tu}$ describe $N\times N$ matrix realization of the algebra (\ref{wz3.2}), i.e. provide the fundamental vectorial realization of the orthogonal rotations generators ${\widetilde{T}}^{rs}$.

{\textbf{a) Nonstandard $N$-extended D=4 Maxwell superalgebras.}

In such a case we supplement the rescalings (\ref{wz2.3}) providing the Maxwell algebra with the following relations
\begin{equation}\label{wz3.4}
{\widetilde{Q}}^{r}_{A}
=
\lambda \, Q^{r}_{A} \qquad
{\widetilde{T}}^{rs} =
\lambda^2 \, {\widetilde{Z}}^{rs} \,.
\end{equation}
Using (\ref{wz2.3}) and (\ref{wz3.4}) we obtain the $N$-extended nonstandard D=4 Maxwell superalgebras
\begin{equation}\label{wz3.5}
\begin{split}
&
\{
Q^r_A , A^s_B 
\}
= 2 \delta^{rs} (C \, \sigma^{\mu\nu})_{AB} Z_{\mu\nu}
+ 
{\widetilde{Z}}^{rs} \, C_{AB}
\\[5pt]
&
[
M_{\mu\nu}, Q^r_A
]
 = \frac{1}{2} (\sigma _{\mu\nu})_{AB} \, Q^r_B
 \\[5pt]
 &
 [
M_{\mu\nu}, Z_{\rho \tau}
]
=
i(
\eta_{\nu[\rho} Z_{\mu]\tau} 
- \eta_{\mu[\rho}Z_{\nu]\tau}
)
\\[5pt]
&
[{\widetilde{Z}}^{rs}, {\widetilde{Z}}^{tu}]
=
[
M_{\mu\nu}, {\widetilde{Z}}^{rs}
] =0
\\[5pt]
&
[
P_\mu , Q^r_A
]
= [Z_{\mu\nu}, Q^r_A ]
= [{\widetilde{Z}}^{rs}, Q^t_A]=0 \,.
\end{split}
\end{equation}
We see that the generators ${\widetilde{Z}}^{rs}$ 
 in (\ref{wz3.5})
 describe the set of 
 $\frac{N(N-1)}{2}$ scalar central charges.
In order to obtain ${\widetilde{Z}}^{rs}$ as the tensorial central charges which transform under internal $O(N)$ as 2-tensors, one should consider the contractions (\ref{wz2.3}) and (\ref{wz3.4}) applied to the following choice of semidirect product (\ref{wz1.7})
\begin{equation}\label{wz3.6}
{\widetilde{G}} = (O(3,1)\oplus O(N)) 
 \ltimes
OSp(N|4)
\end{equation}

{\textbf{b) Standard $k$-extended Maxwell superalgebra.}}

Let us  consider the contraction of the superalgebra (\ref{wz3.1}) for even $N$ ($N=2k$). We introduce the following projected supercharges (see also \cite{mosluk18})
\begin{equation}\label{wz3.7}
{\widetilde{Q}}^{(\pm)r}_{\quad A}
= P^{(\pm)rs}_{(k)\ AB} \, 
{\widetilde{Q}}^{s}_B\,,
\end{equation}
where ($r,s=1,2\ldots 2k; \
\Omega^{rs} = - \Omega^{sr}$, $\Omega^2 =-1$)
\begin{equation}\label{wz3.8}
P^{(\pm)rs}_{(k) \ AB}
=\frac{1}{2}
(
\delta_{AB} \,\delta^{rs} \pm C _{AB} \, \Omega^{rs}
)\,.
\end{equation}
It follows from (\ref{wz3.8}) that 
\begin{equation}\label{wz3.9}
P^{(\pm)rs}_  {\quad AB} \,
P^{(\pm)st}_{\quad BC}
=  P^{(\pm)rt}_{\quad AC}
\qquad
P^{(+)rs}_  {\quad AB} \,
P^{(-)st}_{\quad BC}
= 0 \, .
\end{equation}
We choose further
\begin{equation}\label{wz3.10}
\Omega = \begin{pmatrix}0 & 1_k
\\[5pt]
 -1_k  &0 
\end{pmatrix}
\, .
\end{equation}
The $2k\times 2k$ antisymmetric operator - valued matrix
${\widetilde{T}}^{rs}$ describing $O(2k)$ generators can be divided into two sectors
 ${\widetilde{T}}^{rs}_+ , {\widetilde{T}}^{rs}_{-}$ satisfying the relations

\begin{equation}\label{wz3.11}
{\widetilde{T}}_{\pm} \Omega
= \mp \Omega \, 
{\widetilde{T}}_{\pm}^T
= \pm \Omega \, 
{\widetilde{T}}_{\pm}\,.
\end{equation}
We get explicitly ($A=-A^T, \, C=-C^T , B$ arbitrary)
\begin{equation}\label{wz3.12}
{\widetilde{T}}_{\pm}
=
\begin{pmatrix}A & B
\\[5pt]
-B^T & C
\end{pmatrix}
\qquad
\begin{array}{l}
B = \pm B^T
\\[5pt]
A = \pm C\,.
\end{array}
\end{equation}
From (\ref{wz3.12}) follows that the generators 
${\widetilde{T}}^{rs}_{+}$ describe the algebra 
$ U(k)$ and the generators 
${\widetilde{T}}^{rs}_{-}$ the coset 
$\frac{O(2N)}{U(k)}$.

Using the projected supercharges (\ref{wz3.7}) one can 
rewrite the $OSp(N|4)$ superalgebra as follows:
\begin{equation}\label{wz3.13}
\begin{split}
&\{
{\widetilde{Q}}^{(\pm)r}_A ,
{\widetilde{Q}}^{(\pm)s}_B
\} =
\frac{1}{2}
P^{(\pm)rs}_{AB}
(C\, \gamma^\mu )_{AB} {\cal P}_\mu
+ C_{AB} \, {\widetilde{T}}_{-}^{rs}
\\[5pt]
&\{
{\widetilde{Q}}^{(+)r}_A ,
{\widetilde{Q}}^{(-)s}_B
\}=
\frac{1}{2}
P^{(+)rs}_{\quad AB}
(C\, \gamma^{\mu\nu})_{AB} {\widetilde{M}}_{\mu\nu}
+ C_{AB} \, {\widetilde{T}}_{+}^{rs}\,.
\end{split}
\end{equation}
We supplement (\ref{wz2.3}) with the following set of rescalings

\begin{equation}\label{wz3.14}
\begin{split}
&
{\widetilde{Q}}^{(+)r}_{\quad A}
= \lambda^{1/2} \, 
Q^{(+) r}_{\quad A}
\qquad
{\widetilde{T}}^{ts}_{+} 
= \lambda^2 T^{rs}_{+}
\\[5pt]
&
{\widetilde{Q}}^{(-)r}_{\quad A}
= \lambda^{3/2} \, Q^{(-) r}_{\quad A}
\qquad
{\widetilde{T}}^{rs}_{-} = \lambda\,
 T^{rs}_{-}\,.
\end{split}
\end{equation}
We obtain in the limit $\lambda \to \infty$ the following
contracted superalgebra
\begin{equation}\label{wz3.15}
\begin{split}
& \{
{\widetilde{Q}}^{(+)r}_A ,
{\widetilde{Q}}^{(+)s}_B
\} =
2
P^{(+)rs}_{\quad AB}
(C\, \gamma^{\mu})_{AB} {P}_{\mu}
+ C_{AB} \, {{T}}_{-}^{rs}
\\[5pt]
&\{
{\widetilde{Q}}^{(+)r}_A,
{\widetilde{Q}}^{(-)s}_B
\} = 
2
P^{(-)rs}_{\quad AB}
(C\, \sigma_{\mu\nu})_{AB} {{Z}}_{\mu\nu}
+ C_{AB} \, {{T}}_{+}^{rs}
\\[5pt]
&\{
{\widetilde{Q}}^{(-)r}_A,
{\widetilde{Q}}^{(-)s}_B
\} =0
\,.
\end{split}
\end{equation}
In such a way we get new $k$-extended standard D=4 Maxwell superalgebra with the internal sector $T$ described by the
 sum of two Abelian subalgebras 
 $ T^{rs}_{+}$ ($k^2$ generators) and
  $ T^{rs}_{-}$ ($k(k-1)$ generators) satisfying the  following relations
\begin{equation}\label{wz3.16}
[\widehat{T}_{-}, \widehat{T}_{-}]
 \subset \widehat{T}_{+} \qquad
 [\widehat{T}_{\pm}, \widehat{T}_{\pm}]
 =0 \,.
\end{equation}
We see therefore that the internal symmetry sector 
 $T=(T_+ , T_-)$ has similar algebraic structure as the space-time sector described by Maxwell algebra $(Z_{\mu\nu}, P_\mu)$
  (compare (\ref{wz3.16}) with (\ref{wz2.4d}--\ref{wz2.4e})).
 If we put $k=1$ we obtain 
 simple standard Maxwell superalgebra (see Sect.~2iii)), with single internal symmetry generator  $T_ + = B_5$  (see (\ref{wz2.15}--\ref{wz2.16})).

\section{Concluding Remarks}

We would like to add the following brief  comments:

a) The choice (\ref{wz1.7})  of nonsimple algebra in the form of semidirect product leads to simple homogeneous rescaling rules (\ref{wz1.6}). If we introduce the generators 
$\widehat{j}^{(\pm)}=\widehat{h}^{(1)} \pm \widehat{h}^{(2)}$, where ${\widehat{h}}^{(i)}$ ($i=1,2$)  is the pair of commuting $\widehat{h}$-algebras,  the algebra 
$\widehat{G}$ can be rewritten equivalently as the direct summ of algebras 
 $\widehat{j}{ }^{(+)}$ and ($\widehat{j}{ }^{(-)} \oplus 
 {\widehat{k}}$)
\begin{equation}\label{wz4.1}
\widehat{G} = \widehat{j}^{(+)} \oplus
(\widehat{j}^{(-)} \oplus \widehat{k})\,.
\end{equation}
Such a way of describing the deformed (super)Maxwell algebra which  provides after contraction the (super)Maxwell algebras
was preferred  in \cite{mosluk16,mosluk19}\footnote{See also \cite{mosluk20} where however the 
contractions of $ O(3,1)\oplus OSp(N;4)$ to the standard
 Maxwell superalgebras were not considered.}.

ii) Maxwell symmetries can be interpreted as describing the symmetries of Minkowski space-time filled uniformly with constant EM field.
In such a geometric framework we assume that arbitrary constant values of EM field strength describe new degrees of freedom, which extend the standard relativistic space-time \cite{mosluk10,mosluk11,mosluk12}, \cite{mosluk15,mosluk16}. Recently proposed simple standard Maxwell superalgebra describes the supersymmetries in superspace enlarged by arbitrary constant  components of the $U(1)$ gauge field strength superfield (for details see \cite{mosluk14}).

iii) Addition of new generators to D=4 Poincar\'{e} algebra leads naturally to extended space-time geometry. If we wish however to stay e.g. with the framework of D=4 field theory invariant under the (super) Maxwell symmetries, one should realize these new symmetries on the target space of field values, in analogy with the component formulation of the supersymmetric extension of standard D=4  relativistic field theory. 
These ideas are now under consideration.

\begin{acknowledgments}
The author would like to acknowledge the fruitful collaboration with Sotirios Bonanos, Joaquim Gomis and Kiyoshi Kamimura on the subject of Maxwell algebras and Maxwell superalgebras, and wishes to thank for valuable discussions. He would like also to thank Jos\'{e} A. de Azc\'{a}rraga for valuable remarks.

He acknowledges also the financial support of Polish Ministry of Science and Higher Education (grant NN202 318534) in particular permitting  the participation of the author in the Conference (Moscow, January 2010) celebrating 70-th birthday of Academician Andrei A. Slavnov. To conclude, the author would like to thank the organizers of the Conference, in particular Ira Aref'eva, for warm hospitality in Moscow.
\end{acknowledgments}

\end{document}